\documentclass[conference]{IEEEtran}

\usepackage{cite}

\usepackage[pdftex]{graphicx}

\usepackage{amsmath}
\usepackage{numprint}
\usepackage[round-precision=3,round-mode=figures,scientific-notation=true]{siunitx}

\usepackage{array}

\usepackage{fixltx2e}
\usepackage{stfloats}

\usepackage{url}
\usepackage{tabularx,lipsum}
\usepackage{multirow}

\usepackage{todonotes}

\begin{document}
\title{Benchmarking the Graphulo Processing Framework}

\author{\IEEEauthorblockN{Timothy Weale\IEEEauthorrefmark{2},
Vijay Gadepally\IEEEauthorrefmark{3}\IEEEauthorrefmark{4}*,
Dylan Hutchison$^\circ$ and
Jeremy Kepner\IEEEauthorrefmark{3}\IEEEauthorrefmark{4}$^+$}
\IEEEauthorblockA{\IEEEauthorrefmark{2}Department of Defense, \IEEEauthorrefmark{3}MIT Lincoln Laboratory, \IEEEauthorrefmark{4}MIT Computer Science \& AI Laboratory\\ $^+$MIT Mathematics Department, $^\circ$University of Washington}}


\maketitle

\begin{abstract}
Graph algorithms have wide applicablity to a variety of domains and are often used on massive datasets. Recent standardization efforts such as the GraphBLAS specify a set of key computational kernels that hardware and software developers can adhere to. Graphulo is a processing framework that enables GraphBLAS kernels in the Apache Accumulo database. In our previous work, we have demonstrated a core Graphulo operation called \textit{TableMult} that performs large-scale multiplication operations of database tables. In this article, we present the results of scaling the Graphulo engine to larger problems and scalablity when a greater number of resources is used. Specifically, we present two experiments that demonstrate Graphulo scaling performance is linear with the number of available resources.  The first experiment demonstrates cluster processing rates through Graphulo's TableMult operator on two large graphs, scaled between $2^{17}$ and $2^{19}$ vertices. The second experiment uses TableMult to extract a random set of rows from a large graph ($2^{19}$ nodes) to simulate a cued graph analytic. These benchmarking results are of relevance to Graphulo users who wish to apply Graphulo to their graph problems.
\end{abstract}
\IEEEpeerreviewmaketitle
\section{Introduction}

Large-scale graph analytics are of interest in a variety of domains. For example in an intelligence, surveillance, and reconnaisance (ISR) application, graph analytics may be used for entity matching or track analysis. For social media applications, graph analytics may be used to determine communities or patterns of interest. Recently, an effort known as the GraphBLAS~\cite{kepner2015graphs,bulucc2011combinatorial} (\url{http://graphblas.org/}) has sought to standardize a set of computational kernels and demonstrate their applicability to a wide variety of graph analytics~\cite{Gadepally:2015aa}. The connection between graph processing and linear algebra is well defined in~\cite{Kepner:2011aa,Kepner:2014ab}. The GraphBLAS specification \cite{Mattson:2014aa,Bader:2014aa,Kepner:2014ab} currently consists of five core operations: sparse matrix multiplication, element-wise matrix multiplication, matrix addition, reference, and dereference, along with a small number of additional helper functions.

\let\thefootnote\relax\footnotetext{*Vijay Gadepally is the corresponding author and can be reached at vijayg [at] ll.mit.edu\\
This material is based upon work supported by the Assistant Secretary of Defense for Research and Engineering under Air Force Contract No. FA8721-05-C-0002.  Any opinions, findings and conclusions or recommendations expressed in this material are those of the author(s) and do not necessarily reflect the views of the Assistant Secretary of Defense for Research and Engineering.}

Given that the vast majority of data are stored in databases such as Apache Accumulo, there is also interest in developing processing frameworks, such as GraphX~\cite{xin2013graphx}, Pregel~\cite{malewicz2010pregel}, Gaffer\cite{gaffer:2016}, and GraphLab~\cite{low2012distributed}, optimized for graph operations on data in databases. A survey of popular graph database models is provided in~\cite{angles2008survey}. It is natural that a processing framework designed for databases makes use of the GraphBLAS kernels.

Graphulo~\cite{Hutchison:2015aa} is a specialized graph processing framework built to work with Apache Accumulo and to conform to the GraphBLAS standard. Apache Accumulo is a NoSQL database designed for high performance ingest and scans\cite{Sen:2013aa, Kepner:2014aa}. As a general-purpose warehousing solution, Accumulo has wide adoption in a variety of government and non-government settings.

\begin{figure}[h]
\begin{center}
\includegraphics[width=1.0\linewidth]{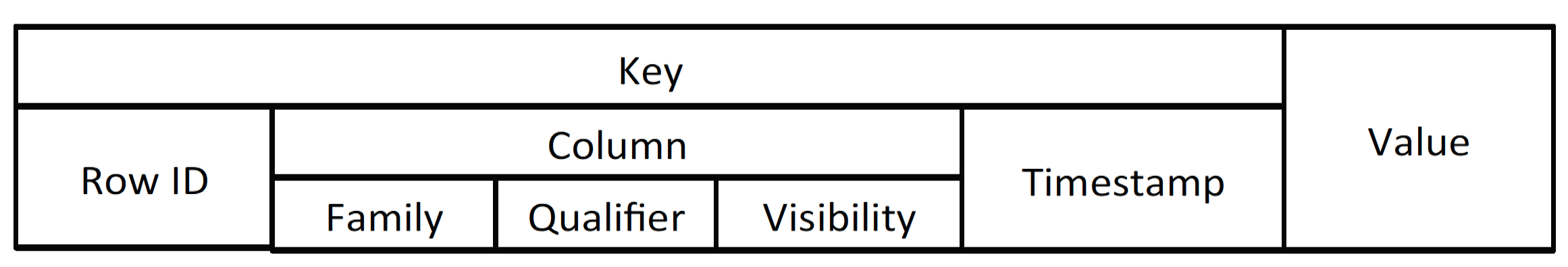}
\caption{Accumulo Key-Value Data Model}
\label{figure:data model}
\end{center}
\end{figure}

As shown in Figure \ref{figure:data model}, Accumulo's data model has a key with 5 components: row, column family, column qualifier, visibility, and timestamp. For simplicity, Graphulo provides directed graph structures using the row (source vertex), column qualifier (destination vertex), and the value (the number of connections between the two vertices). The visibility and timestamp attributes are not investigated in this paper.

One of Accumulo's advantages over other data warehousing solutions is its \textit{iterator} framework for in-database computation. This capability provides improved performance by exploiting data locality, avoiding unnecessary network transfers, and moving the processing to the stored data. Additionally, by co-locating storage and computation, one can take full advantage of other aspects of Accumulo's infrastructure, such as write-ahead logging, fault-tolerant execution, and parallel load balancing of data. As a first-order operator, iterators have full access to Accumulo's data model, allowing for highly flexible, server-side algorithm development capabilities.


In the remainder of this paper, we provide a brief high-level review of Graphulo in Section \ref{sec:Graphulo}, including a description of the Accumulo iterators used and the process for tracking the amount of work done in each experiment. Section \ref{sec:Experiment Graph Multiplication} presents the main SpGEMM scaling experiment using Graphulo, including experimental setup and results. We then describe a row extraction experiment in Section \ref{sec:Experiment Sub-Graph Extraction}, providing scaling results for another real-world Graphulo use case. Finally, we discuss related work and areas for optimization in Section \ref{sec:Discussion and Related Work} before concluding in Section \ref{sec:Conclusion}.

\section{Graphulo}\label{sec:Graphulo}

At a high level, the goal of the Graphulo project is to allow developers to focus on analytic development, rather than on writing custom MapReduce processes for graph data stored in Apache Accumulo. Thus, Graphulo explicitly implements the GraphBLAS specification on top of Apache Accumulo\cite{Gadepally:2015aa}. In our experiments, we focus on testing the performance of the Sparse Generalized Matrix Multiply (SpGEMM), the core kernel at the heart of GraphBLAS, because SpGEMM can be readily extended (via user-defined multiplication and addition functions) to express many GraphBLAS primitives. Additionally, SpGEMM is an essential part of a wide range of algorithms, including graph search, table joins, and many others \cite{Kepner:2011aa}.

Given matrices \textbf{A} and \textbf{B} as well as operations $\oplus$ and $\otimes$ for scalar addition and multiplication,
the matrix product $\mathbf{C} = \mathbf{A} {\,\oplus.\otimes\,} \mathbf{B}$, or for convenience, \textbf{C = AB}, defines entries of result matrix \textbf{C} as

\begin{equation}
\mathbf{C}(i, j) = \bigoplus_{k} \mathbf{A}(i, k) \otimes \mathbf{B}(k, j)
\label{equation:tablemult}
\end{equation}

\begin{figure}[htb]
\begin{center}
\includegraphics[width=0.9\linewidth]{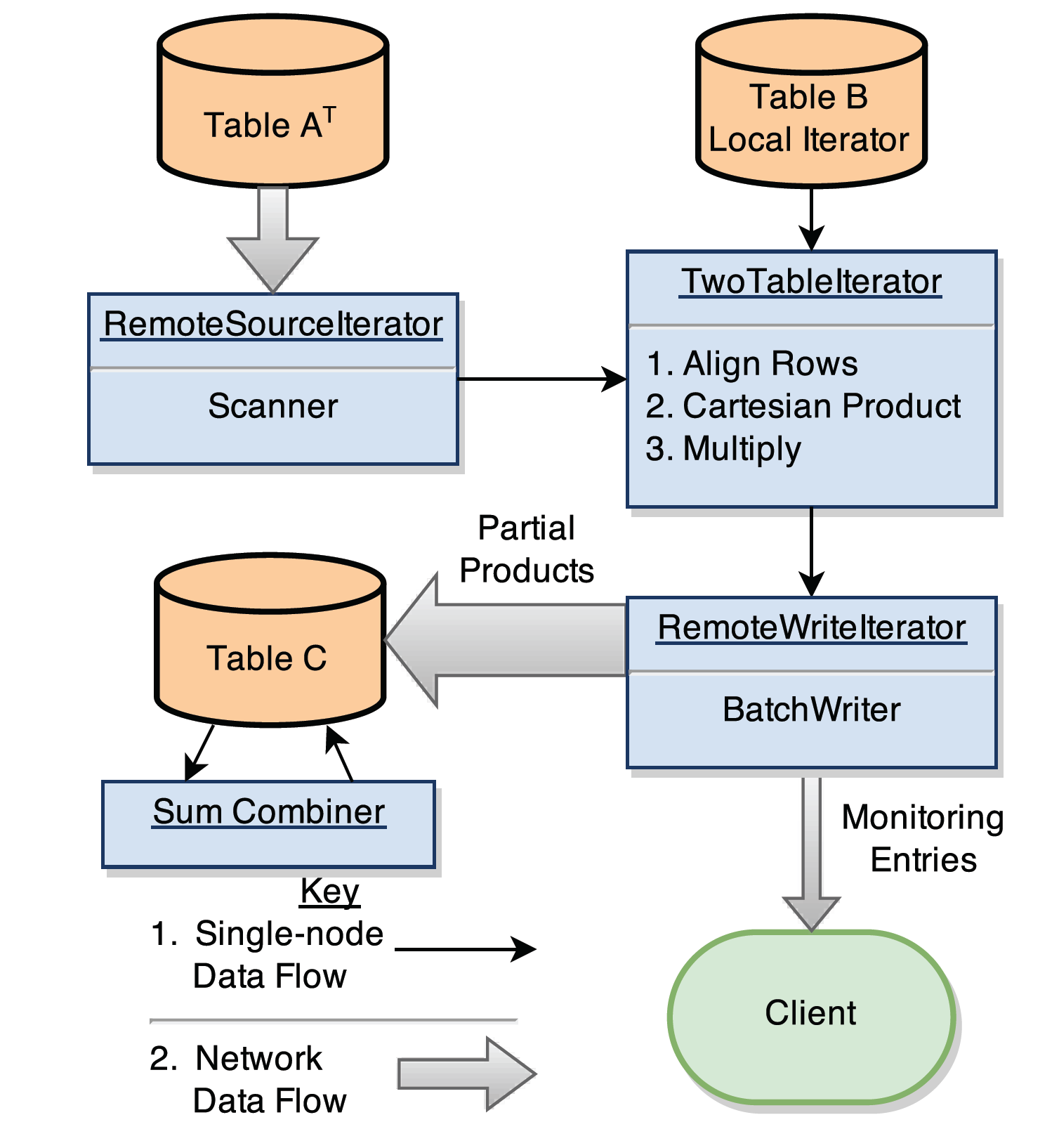}
\caption{Graphulo TableMult Iterator Processing Steps and Data Flow}
\label{figure:tableMult}
\end{center}
\end{figure}

We call the implementation of SpGEMM in Accumulo \textit{TableMult}, short for multiplication of Accumulo tables. Accumulo tables have many similarities to sparse matrices, although a more precise mathematical definition is associative arrays\cite{kepner2015associative}. For this work, we concentrate on distributed tables that may not fit in memory and use a streaming approach that leverages Accumulo's built-in distributed infrastructure. In Figure \ref{figure:tableMult}, we present the Graphulo \textit{TableMult} iterator stack that implements Equation \ref{equation:tablemult}. 
TableMult operates on a table storing the transpose $\mathbf{A}^T$ 
and a table storing $\mathbf{B}$ and writes its output to a separate table $\mathbf{C}$.

TableMult's execution involves three Accumulo iterators: the \emph{RemoteSourceIterator}, the \emph{TwoTableIterator} and the \emph{RemoteWriteIterator}. The RemoteSourceIterator scans the entries of table $\mathbf{A}^{T}$ and provides them as inputs to the TwoTableIterator, which aligns and creates partial products. This action is the implementation of the $\mathbf{A}(i, k) \otimes \mathbf{B}(k, j)$ portion of Equation \ref{equation:tablemult}. We call intermediary results of operations \textit{partial products (pp)}. For the sake of sparse matrices, we only perform $\oplus$ and $\otimes$ when both operands are nonzero. The processing rate of the Graphulo cluster is given by the number of partial products divided by the processing time.

The results of the multiplication are then written to table $\mathbf{C}$ with the RemoteWriteIterator. The summation ($\oplus$ operator) is accomplished through combiner iterator subclasses placed on $\mathbf{C}$. This lazy summation only runs when table $\mathbf{C}$ is required to execute a scan or minor/major compaction and helps maximize throughput while guaranteeing correctness of output.

\section{Experiment: Graph Multiplication}\label{sec:Experiment Graph Multiplication}

\begin{table*}[t]                                                         
\caption{Numerical Results and Parameters for Large-Graph \textit{TableMult} Multiplication Experiments}
\label{table:exp1_table}                                              
\centering                                                                
\begin{tabular}{|c||c|c|c|c||c|c|c|c|c|}
\hline
Nodes & SCALE & Partial Products (pp) & Rate (pp/sec) & Speedup Ratio & SCALE & Partial Products (pp) & Rate (pp/sec) & Speedup Ratio\\
\hline
1 & \multirow{5}{*}{18x18} & \multirow{5}{*}{\num{2935400e3}} & \num{192460.0052} & 1.00 & 17x17 & \num{1090300000} & \num{182642.0531} & 1.00 \\   
\cline{1-1}\cline{4-9}
2 & & & \num{386011.3679} & \num{2.01} & 18x17 & \num{1683200000} & \num{400659.2888} & \num{2.193685857} \\   
\cline{1-1}\cline{4-9}
4 & & & \num{752782.4794} & \num{3.91137098} & 18x18 & \num{2935400e3} & \num{761432.9071} & \num{4.168990078}\\   
\cline{1-1}\cline{4-9}
8 & & & \num{1380779.905} & \num{7.174373207} & 19x18 & \num{4527300e3} & \num{1364178.745} & \num{7.46913825}\\  
\cline{1-1}\cline{4-9}
16 & & & \num{1874217.852} & \num{9.738219895} & 19x19 & \num{7847500e3} & \num{1900121.065} & 10.40\\  
\hline
\end{tabular}                                                             
\end{table*}

Prior work~\cite{Hutchison:2015aa} compared Graphulo's performance against the Dynamic Distributed Dimensional Data model (D4M), demonstrating improvement in terms of both speed and scale of computation. In this experiment, we demonstrate Graphulo's ability to scale its processing with the number of cluster nodes. As in the previous experiment, we create and store random input graphs in our Accumulo tables using the Graph500 unpermuted power-law graph generator~\cite{Bader:2006aa}. Power-law graphs have a highly connected initial vertex and exponentially decreasing degrees for subsequent vertices and have been used to describe a diverse set of real graphs~\cite{gadepally2015using}. The power-law generator takes \emph{SCALE} and \emph{EdgesPerVertex} as parameters, creating graphs with $2^{SCALE}$ rows and $EdgesPerVertex \times 2^{SCALE}$ entries. For our experiments, we fix EdgesPerVertex to $16$ and use the SCALE parameter to vary problem size. 

Our experimental setup is as follows:

\begin{enumerate}
\item Generate two graphs with different random seeds and insert them into Accumulo as adjacency tables.
\item For each table, identify optimal split points for each input graph. These optimal split points depend on the number of tablets,$^*$\footnotemark[1] which scales with the number of nodes in the cluster.\footnotetext[1]{$^*$In our configuration, a single-node cluster had each table split across two tablets; two nodes had four tablets; etc.}
\item Set the input graphs' table splits equal to that point.
\item Create an empty output table. Pre-split it with an input split position from one of the two input tables.
\item Compact the input and output tables so that Accumulo redistributes the tables' entries into the assigned tablets.
\item Run and time Graphulo TableMult multiplying the transpose of the first input table with the second.
\item Repeat the experiment several times to derive a representative processing rate and minimize noise.
\end{enumerate}

Previous work included an additional processing step to determine the optimal tablet split for the output graph. We include no such ``oracle'' capability and set the output node distribution on the basis of one of the two input graphs, which can be calculated at ingest. As the number of nodes and tablets increased, calculating the optimal split became an overwhelming part of the experimental runtime.

The experiments were conducted on a shared system, with each node initially allocated 4GB of memory to an Accumulo tablet server  (allowing growth in 1G steps), 1GB for Java-managed memory maps and 128MB for each data and index cache. The installed software was Graphulo 1.0.0, Accumulo 1.6.0, Hadoop 2.2.0 and ZooKeeper 3.4.6. Each table was split across two tablets per node, allowing for some additional processing parallelization.

\begin{figure}[h]
\begin{center}
\includegraphics[width=1.0\linewidth]{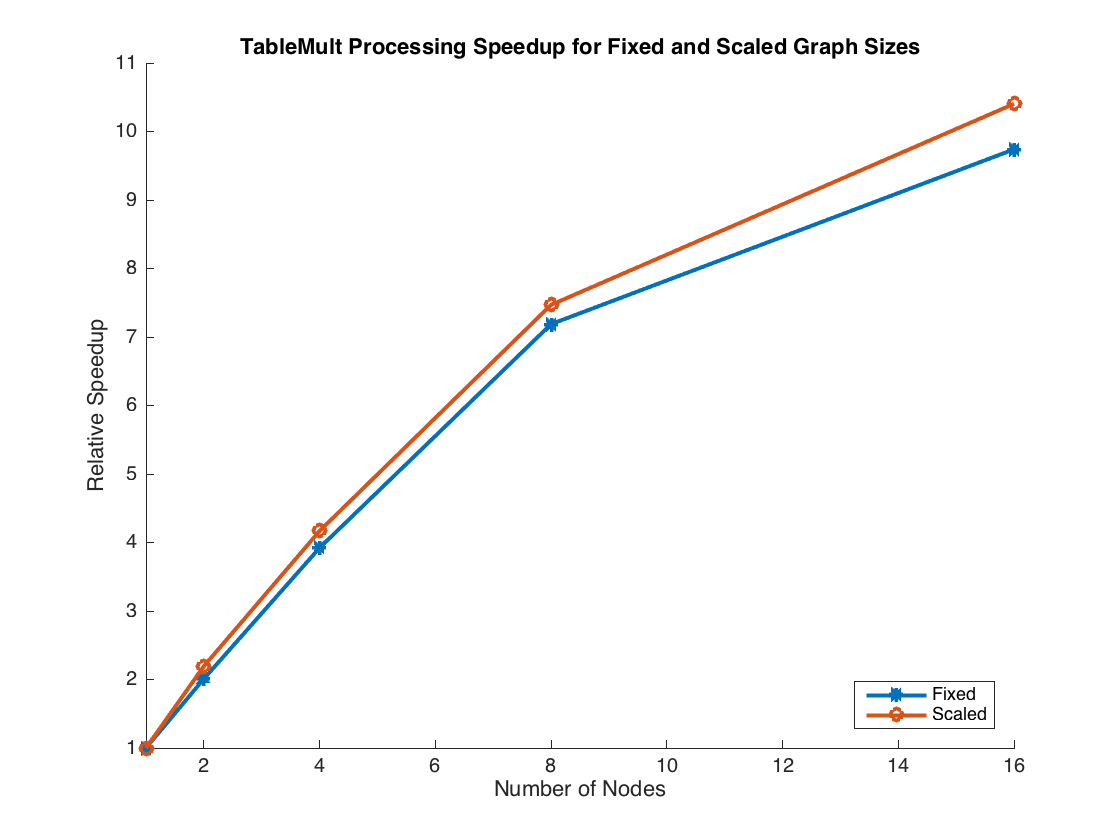}
\caption{Relative Processing Speedup for TableMult Operation Across Fixed-Size and Scaled Graphs}
\label{figure:exp1_speedup}
\end{center}
\end{figure}

Figure \ref{figure:exp1_speedup} shows the results for the first experiment. In this, the single-node cluster processing rate is used as a baseline for comparison. All other plots demonstrate the relative speedup from the single-node example. Ideal linear speedup is a line corresponding to a plot of $y=x$.

In the ``Fixed'' experiment, the size of the graphs involved in the TableMult operation were held constant at SCALE = 18. This strong-scale experiment allowed us to examine the processing speedup across the entire cluster by standardizing the total amount of work to be done and normalizing the entire cluster performance against a single node. The ``Scaled'' experiment increases the size of one of the graphs in the TableMult operation with the number of nodes. Increasing both the graph size and number of nodes by a power of 2 allows for a roughly linear growth pattern that keeps the amount of work done per node roughly constant, providing weak scaling results for Graphulo.

The experiment demonstrates that Graphulo retains Accumulo's ability to grow linearly with the number of processing nodes. In both the strong and weak scaling experiments, Graphulo closely mirrored the expected linear speedup. Numerical results of the first experiment can be found in Table \ref{table:exp1_table}. Some tweaks (discussed below) should improve performance in the future.

\section{Experiment: Subgraph Extraction}\label{sec:Experiment Sub-Graph Extraction}

Organizations often have a massive amount of information available in their databases. At any time, however, only certain subsets of interest will be required for analysis and processing. Graphulo can provide a quick way to extract and use these subsets, improving an algorithm's ability to run effectively. We refer to such analytics as cued analytics, that is, analytics on selected table subsets. Such analytics are ideally suited for databases by quickly accessing subsets of interest (for example, particular communities in a large social media graph) in contrast to whole-table analytics that may be better suited for parallel processing frameworks such as Hadoop Map-Reduce~\cite{Burkhardt:2015aa}.

In this experiment, we take a large sample graph (SCALE = 19) and randomly extract some portion for analysis (in our experiments, either 1024 or 2048 graph edges) by randomly generating a set of vertices to sample ($\{sampleset\}$). This random set of vertices is then used to create a binary diagonal matrix (using Equation \ref{Extraction Graph}) that contains $1$s at the randomly selected vertex locations (Equation \ref{Extraction Graph Logic}). These operations are illustrated below:

\begin{equation}\label{Extraction Graph}
E = 
 \begin{pmatrix}
  a_{1,1} & 0 & \cdots & 0 \\
  0 & a_{2,2} & \cdots & 0 \\
  \vdots  & \vdots  & \ddots & \vdots  \\
  0 & 0 & \cdots & a_{n,n} 
 \end{pmatrix}
\end{equation}

\begin{equation}\label{Extraction Graph Logic}
a_{n,n}= 
\begin{cases}
    1 & \text{if } n \in \{sample set\}\\
    0 & \text{otherwise}
\end{cases}
\end{equation}

The generated binary diagonal matrix ($E$) can then be used to extract the rows ($G_{samp}$) from the original graph ($G_{orig}$) by using a TableMult operation with the two graphs ($E$,$G_{orig}$). The experimental steps are as follows:



\begin{enumerate}
\item Insert a single, large graph into Accumulo as adjacency tables. This will be the graph for the entire experiment.
\item Generate a random diagonal matrix with the appropriate number of non zero entries (1024, 2048).
\item Extract a set of random rows using the Graphulo TableMult operation.
\item Repeat the random extraction process several times to derive a representative processing rate.
\end{enumerate}

We use the same cluster configuration as in the prior experiment: a shared system with each node initially allocated 4GB of memory to an Accumulo tablet server  (allowing growth in 1G steps), 1GB for Java-managed memory maps and 128MB for each data and index cache. The installed software was Graphulo 1.0.0, Accumulo 1.6.0, Hadoop 2.2.0 and ZooKeeper 3.4.6. Each table is split across 2 tablets per node to allow for additional processing parallelization.

As mentioned in Section \ref{sec:Graphulo} and shown in Figure \ref{figure:tableMult}, Graph $A$ is the ``remote'' graph in the TableMult implementation. We see significant speedup when the extraction graph ($E$) is used in the first position, because 
fewer entries are sent over the network than when the $E$ and $G_{orig}$ switch positions.

\begin{figure}[h]
\begin{center}
\includegraphics[width=0.99\linewidth]{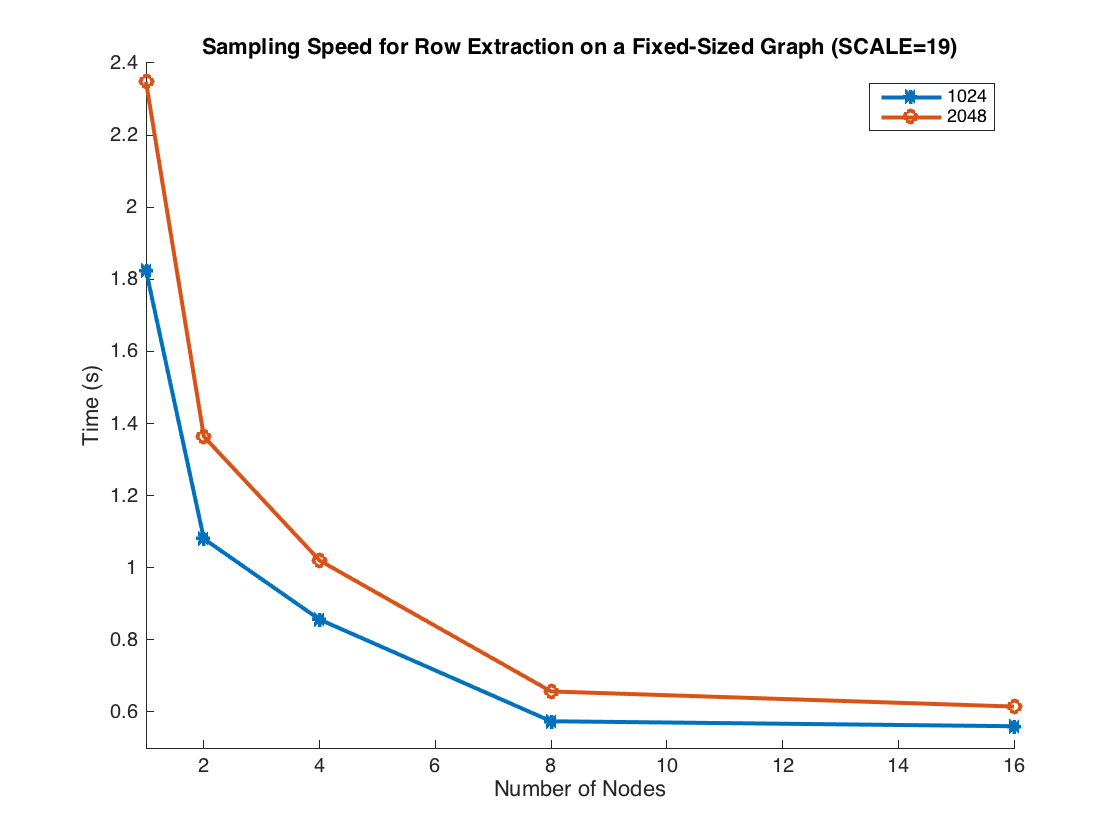}
\caption{Graphulo Row Extraction Experiment Results for Fixed Total Problem Size (Strong Scaling)}
\label{figure:exp2_fixed}
\end{center}
\end{figure}

Figure \ref{figure:exp2_fixed} presents the results of our strong scaling experiment, where the graph size is SCALE = 19 and the number of extracted rows was 1024 or 2048 is kept fixed and we vary the number of processing elements. This graph clearly shows the expected drop-off in the time required to extract the number of rows as we increase the number of nodes. The sample times are within an acceptable range for interactive queries on a large dataset.

\begin{figure}[t]
\begin{center}
\includegraphics[width=0.99\linewidth]{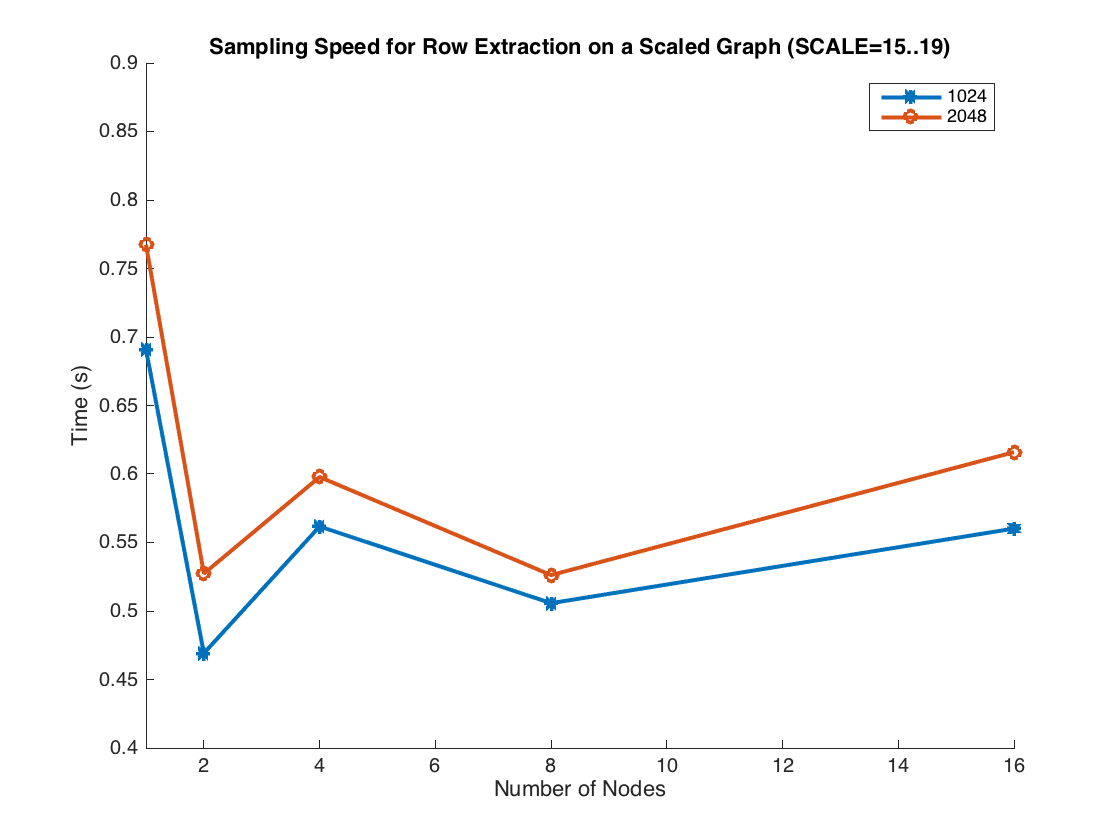}
\caption{Graphulo Row Extraction Experiment Results for Fixed Problem Size Per Node (Weak Scaling)}
\label{figure:exp2_scaled}
\end{center}
\end{figure}

Figure \ref{figure:exp2_scaled} presents the results of our weak scaling experiment, where the graph SCALE parameter changes from 15 to 19 (depending on the number of nodes) and the number of extracted rows is fixed to either 1024 or 2048. These results also indicate the expected constant-time performance as the amount of work per node is held constant. As before, we see a performance difference of approximately $10\%$ in computation time between the experiment on rows of 1024 and 2048.

These results demonstrate that Graphulo can provide a quick and effective method of extracting rows from a larger in-database graph for analysis. The extracted cued graph exists in the Accumulo database as a new table, available for additional local processing for analysis without anyone having to copy large amounts of data across the network or set up filtered iterators on the large graph. This capability enables the size of the cued graph data, not the size of the dataset from which it was sampled, to be the primary factor in Graphulo analytics.

\begin{table*}[t]                                                         
\caption{Numerical Parameters and Results for the Graphulo Row Extraction Experiments}
\label{table:Experiment 2}                                              
\centering                                                                
\begin{tabular}{|c||c|c|c|c|c||c|c|c|c|c|}
\hline
Nodes & SCALE & \# Rows & Partial Prod & Seconds &  Speedup & SCALE & \# Rows & Partial Prods & Seconds & Speedup\\
\hline
\multirow{2}{*}{1} & \multirow{2}{*}{19} & 1024 & \num{13180} & \num{1.8253} & \num{1.0} & \multirow{2}{*}{15} & 1024 & \num{15061.2} & 0.69 & \num{1.0}\\
& & 2048 & \num{26680} & \num{2.34775} & \num{1.0} & & 2048 & \num{31275.9} & 0.77 & \num{1.0}\\
\hline
\multirow{2}{*}{2} & \multirow{2}{*}{19} & 1024 & \num{14060.5} & \num{1.0803} & \num{1.799018519} & \multirow{2}{*}{16} & 1024 & \num{14777.1} & 0.47 & \num{1.431117228}\\
& & 2048 & \num{30700.66667} & \num{1.3638} & \num{1.971830887} & & 2048 & \num{29755.2} & 0.53 & \num{1.37849519}\\
\hline
\multirow{2}{*}{4} & \multirow{2}{*}{19} & 1024 & \num{15385.6} & 0.86 & \num{2.487073609} & \multirow{2}{*}{17} & 1024 & \num{15387.15} & 0.56 & \num{1.241132012}\\   
& & 2048 & \num{30688.9} & \num{1.0200335} & \num{2.629003692} & & 2048 & \num{30199.47368} & 0.60 & \num{1.235755796}\\
\hline
\multirow{2}{*}{8} & \multirow{2}{*}{19} & 1024 & \num{15385.6} & 0.57 & \num{3.709599909} & \multirow{2}{*}{18} & 1024 & \num{15354.7} & 0.51 & \num{1.392213831} \\  
& & 2048 & \num{30688.9} & 0.66 & \num{4.10563785} & & 2048 & \num{30918.2} & 0.53 & \num{1.438086396}\\
\hline
\multirow{2}{*}{16} & \multirow{2}{*}{19} & 1024 & \num{15604.6} & 0.56 & \num{3.883833478} & \multirow{2}{*}{19} & 1024 & \num{15604.6} & 0.56 & \num{1.264567818} \\  
& & 2048 & \num{30199.57895} & 0.61 & \num{4.288626495} &  & 2048 & \num{30688.9} & 0.62 & \num{1.221847368}\\
\hline
\end{tabular}                                                             
\end{table*}


\section{Discussion}\label{sec:Discussion and Related Work}

Accumulo has many configuration parameters available for tweaking. The authors focused on scaling characteristics of Graphulo and provided a minimum amount of customization for our particular setup. A fully optimized Accumulo cluster should provide additional performance. However, we believe that the underlying scaling capabilities of Graphulo demonstrated above would only improve from additional tweaking.

Previous experiments used Accumulo's ability to leverage the operating system's native memory maps. Due to configuration management considerations, this option was not available on the shared cluster. Therefore, the above experiments were run using Java's built-in memory management. Informal experimentation has demonstrated a 2x speedup in performance rate between the two options. Additional stability is possible by not relying on Java's garbage collection system, which can sometimes result in timeouts and failed nodes across the entire cluster. Finally, the stability provided by native memory maps should provide opportunities for additional parallelization by enabling additional tablet splits per table.


At this time, column families, visibility labels, and timestamps are not taken into account in our Graphulo experiments. This work is potentially useful and interesting, especially for subgraph-based computations that may involve comparisons over data access policies, different time periods, or different types of graph information.

\section{Conclusions}\label{sec:Conclusion}

In this paper, we presented experiments that demonstrated the scaling properties of Graphulo as its underlying Accumulo cluster grows from one node to sixteen nodes. Our first experiment was a large-scale graph multiplication task that extended previously published work. The second experiment demonstrated Graphulo's ability to quickly extract a set of vertices from a larger graph, which is useful in a more common use case where only a subset of a large stored graph is needed for analysis. In both experiments, we were able to demonstrate the expected strong and weak scaling in our processing rate. Additionally, we presented several areas of additional Graphulo improvement and experimentation.

For future work, we are interested in developing a greater suite of algorithms that can be directly developed using Graphulo such as dimensional analysis~\cite{Gadepally:2014aa} and big data sampling~\cite{gadepally2015sampling}. Further, we would like to investigate providing a Pig interface to Graphulo as well as calling Graphulo operations from the Apache Spark framework.

\section*{Acknowledgments}
The authors would like to thank the Lincoln Laboratory Supercomputing Center team members: Bill Arcand, Bill Bergeron, David Bestor, Chansup Byun, Mike Houle, Matt Hubbell, Mike Jones, Anna Klein, Pete Michaleas, Lauren Milechin, Julie Mullen, Andy Prout, Tony Rosa, Siddharth Samsi, Chuck Yee, and Albert Reuther.



\footnotesize
\bibliographystyle{IEEEtran}
%
\bibliography{Weale}



\end{document}